\documentclass[fleqn,12pt]{article}
\usepackage{amsfonts,epsfig}

\newcommand{\Ga}{\alpha}
\newcommand{\Gb}{\beta}

\newcommand{\Gd}{\delta}
\newcommand{\Geps}{\epsilon}
\newcommand{\Ge}{\varepsilon}
\newcommand{\Gg}{\gamma}
\newcommand{\GG}{\Gamma}

\newcommand{\Gl}{\lambda}

\newcommand{\Go}{\omega}

\newcommand{\Gth}{\theta}

\newcommand{\BGa}{{\bar{\alpha}}}
\newcommand{\BGb}{{\bar{\beta}}}

\newcommand{\CD}{{\cal D}}

\newcommand{\CL}{{\cal L}}
\newcommand{\CO}{{\cal O}}

\newcommand{\bbZ}{{\mathbb{Z}}}

\newcommand{\Beps}{\overline{\Geps}}

\newcommand{\Bpsi}{\overline{\psi}}
\newcommand{\Blambda}{\overline{\lambda}{}}
\newcommand{\BGl}{\overline{\lambda}}

\newcommand{\Bphi}{{\overline{\phi}{}}}

\newcommand{\BF}{\overline{F}}
\newcommand{\Bf}{\overline{f}}
\newcommand{\BQ}{\overline{Q}}

\newcommand{\Tr}{{\tilde r}}

\newcommand{\ft}[2]{{\textstyle {\frac{#1}{#2}} }}

\newcommand{\dd}{\partial}

\newcommand{\ra}{\rightarrow}
\newcommand{\I}{{ i}}

\newcommand{\be}{\begin{equation}}
\newcommand{\ee}{\end{equation}}
\newcommand{\ben}{\begin{displaymath}}
\newcommand{\een}{\end{displaymath}}
\newcommand{\ba}{\begin{eqnarray}}
\newcommand{\ea}{\end{eqnarray}}
\newcommand{\nn}{\nonumber}
\newcommand{\non}{\nonumber\\}

\newcommand{\mathon}{\mathversion{bold}}
\newcommand{\mathoff}{\mathversion{normal}}

\newcommand{\la}{\label}

\newcommand{\Ref}[1]{(\ref{#1})}

\newcommand{\equ}{\!=\!}

\newcommand{\num}{p}

\newcommand{\shc}{\sinh\!\chi}
\newcommand{\chc}{\cosh\!\chi}
\newcommand{\thc}{\tanh\!\chi}

\makeatletter
\@addtoreset{equation}{section}
\makeatother

\begin{document}

\begin{titlepage}

\begin{flushright}  
 
HUB-EP-01/56 , \ \ IHES/P/01/52 , \ \ SPIN-2001/30 , 
\ \ Imperial/TP/01-2/6, \ \ 
ESI-1106 \\

\end{flushright}  
  
\vspace{.6cm}  
\begin{center}  
  
{\Large \bf{Chern-Simons Vortices in Supergravity}}  
  
\vspace{1cm}  
  
M.\ Abou-Zeid\footnote{m.abouzeid@ic.ac.uk} and
H.\ Samtleben\footnote{H.Samtleben@phys.uu.nl} 

\vspace{.5cm}  
$^1${\em Theoretical Physics Group, The Blackett Laboratory,\\ Imperial
College of Science, Technology and Medicine,\\ 
Prince Consort Road, London SW7 2BZ, United Kingdom}

\vspace{.5cm}

$^1${\em Institut des Hautes Etudes Scientifiques,\\ Le Bois-Marie, 35 route
de Chartres,  91440
Bures-sur-Yvette,  France}

\vspace{.5cm}

$^2${\em Spinoza Insituut, Universiteit Utrecht,\\ Leuvenlaan 4,
Postbus 80.195, 3508 TD Utrecht, The Netherlands }

\vspace{1cm}  
\begin{abstract}

We study supersymmetric vortex solutions in three-dimensional abelian
gauged supergravity. First, we construct the general $U(1)$-gauged
$D\!=\!3$, $N\!=\!2$ supergravity whose scalar sector is an arbitrary
K\"ahler manifold with $U(1)$ isometry. This construction clarifies
the connection between local supersymmetry and the specific forms of
some scalar potentials previously found in the literature --- in
particular, it provides the locally supersymmetric embedding of the
abelian Chern-Simons Higgs model.  We show that the Killing spinor
equations admit rotationally symmetric vortex solutions with
asymptotically conical geometry which preserve half of the
supersymmetry.

\end{abstract}
  
PACS numbers: 04.65.+e, 11.27.+d


\end{center}  
\vspace{1.3cm}
 
\flushleft{December, 2001}
 
\end{titlepage}

\vspace*{-4ex}

\section{Introduction}

In three space-time dimensions, pure Einstein gravity has no local
propagating degrees of freedom and is thus topological. The solutions
to the field equations are locally flat except at conical
singularities at the location of matter
sources~\cite{DeJaHo84,AbGiKu84}. Moreover, there is a precise sense
in which the theory is soluble~\cite{Witt88}--\cite{Mats99}. Similar
results hold for cosmological Einstein gravity~\cite{DesJac84}, and
for the three-dimensional topological supergravity theories
\cite{DesKay83,AchTow86}.

A complete classification of $N$-extended supergravities in three
dimensions was given in~\cite{dWToNi93}. In particular, the geometry
of the target manifolds parametrized by the scalar fields is
K\"{a}hler for $N=2$, quaternionic for $N=3,4$ and symmetric for
$N=5,6,8$. For $N=9,10,12,16$, the theories are based on a single
supermultiplet and are associated with coset spaces with the
exceptional isometry groups $F_4$, $E_6$, $E_7$, and $E_8$,
respectively. Recently, a number of maximal ($N\equ16$) gauged models
with a variety of admissible compact and noncompact gauge groups were
constructed in~\cite{NicSam01,NicSam01a}. In these three-dimensional
gauged supergravities, a key role is played by the on-shell duality
between the gauge fields and the scalar fields. This is implemented in
the lagrangian by means of a Chern-Simons term for the gauge fields
(rather than the usual Yang-Mills term), which insures that the
duality relation is an equation of motion while the gauge fields do
not carry physical degrees of freedom.


In this paper we construct the general $U(1)$-gauged non-linear sigma
model coupled to $N=2$ supergravity in three dimensions, and study
supersymmetric vortex solutions of this theory. As particular
examples, we obtain the supersymmetric embedding of the abelian
Chern-Simons Higgs model coupled to
gravity~\cite{Valt92}--\nocite{CanLee92,Lond95,Clem96}\cite{CCKK01}
and generalizations of the $CP^n$ and $CH^n$ models, recently
constructed in~\cite{DKSS00}.

The abelian Higgs model with a Chern-Simons term in three-dimensional
flat Minkowski space and its vortex solutions were studied
in~\cite{HoKiPa90,JacWei90} (see also the earlier references
therein). This model is of some practical interest because of its
relation to the physics of high temperature superconductors, which
violate both the T and P symmetries (like the Chern-Simons term) and
which often exhibit two-dimensional spatial structures. In particular,
it was found~\cite{HoKiPa90,JacWei90} that the model with a specific
sixth-order Higgs potential admits topologically stable vortex
solutions which satisfy (first order) self-duality, or
Bogomol'nyi-type~\cite{Bogo76} equations. This special Higgs potential
has a $U(1)$-symmetric minimum which is degenerate with a
symmetry-breaking one, as a result of which it also admits charged
nontopological soliton solutions~\cite{JaLeWe90}. In
ref.~\cite{LeLeWe90}, the specific form of this potential was shown to
originate from the unique embedding of this model into an globally
$N=2$ supersymmetric theory; the Bogomol'nyi bound may be obtained
from the superalgebra and is saturated by the supersymmetric
solutions.


In the present paper --- as a byproduct of our general construction
--- we give a similar explanation for the origin of the specific {\em
eighth-order} potential found in the abelian Chern-Simons-Higgs model
coupled to gravity. More precisely, it has been found
in~\cite{Valt92,CanLee92,Lond95} that the Einstein equations and the
matter field equations of this model can be recast into a set of
self-duality equations for a specific eighth-order choice of the Higgs
potential which reduces to the sixth-order potential of the flat space
model when the Newton gravitational coupling constant is set to
zero. We show that this is the unique potential which may be embedded
into a locally supersymmetric theory, with the Bogomol'nyi-type
equations descending from the Killing spinor equations of this
underlying supergravity. This in particular allows us to address the
stability of the vortex solutions studied
in~\cite{Valt92}--\cite{CCKK01}.

In the second part of the paper we study supersymmetric vortex
solutions of the general $U(1)$-gauged supergravity with a single
complex scalar field. We show that with a rotationally symmetric
ansatz, the Killing spinor equations reduce to a set of four
first-order differential equations. This generalizes the results
of~\cite{Lond95,CCKK01}, where these equations were found by the
Bogomol'nyi-type arguments mentioned above, to arbitrary K\"ahler
manifolds. Identifying an integral of motion, we show that after
suitable redefinition of coordinates, these equations may further be
reduced to a single second-order differential equation. For a given
vortex number, it has a unique solution with regular asymptotics, from
which all the original fields may be restored. It represents a
rotationally symmetric, finite energy, topologically stable vortex
solution, preserving one half of the supersymmetry.

Another motivation for the study of locally supersymmetric theories in
three dimensions and their solutions stems from an observation made by
Witten~\cite{Witt95}: in three dimensions, the cosmological constant
of the vacuum can be exactly zero because of local supersymmetry, yet
the spectrum of excited states may not exhibit the usual Bose-Fermi
mass degeneracy because for non-zero energy states the supercharges
are defined in conical space-times.  A realization of this mechanism
in the $N=2$ supersymmetric abelian Maxwell-Higgs model coupled to
gauged three-dimensional supergravity was exhibited
in~\cite{BeBeSt95}, and studied further e.~g.\ in
refs.~\cite{EdNuSc96,EdNuSc96b,Edel97}. The abelian gauged $N=2$
supergravities and their half supersymmetric vortex solutions
constructed in this paper provide additional examples of this
mechanism. As in~\cite{BeBeSt95,EdNuSc96b}, the covariantly constant
spinors required to define these supersymmetries exist by virtue of a
cancellation between the Aharonov-Bohm phase and the phase associated
with the holonomy of the spin connection. However, the same mechanism
as in~\cite{BeBeSt95} prevents the existence of normalizable
covariantly constant spinors associated with the other half of the
supersymmetry transformations, and hence Bose-Fermi degeneracy is
absent in the soliton spectrum. Of course, it remains to be seen
whether a locally supersymmetric four-dimensional theory with zero
cosmological constant but without the phenomenologically unviable
Bose-Fermi degeneracies can be constructed along the lines suggested
in~\cite{Witt95}.

The plan of this paper is as follows. In Section~\ref{sec_gauging}, we
construct the general $U(1)$-gauged $N\equ2$ supergravity by deforming
the three-dimensional sigma model of ref.~\cite{dWToNi93}, whose
target space is an arbitrary K\"ahler manifold with $U(1)$
isometry. We show that our results reduce to those of~\cite{DKSS00}
when the target spaces are the homogeneous spaces $CP^n$ and $CH^n$
(and moreover include a possible extension by Fayet-Iliopoulos terms),
while for the complex plane they give the supersymmetric embedding of
the abelian Chern-Simons-Higgs model coupled to gravity. In
Section~\ref{sec_susy}, we study the Killing spinor equations of the
general model with a single complex scalar field and we find that they
reduce to a set of four first-order differential equations. We then
show that they admit a unique solution with the prescribed (regular)
asymptotics and give some numerical examples. We close with a summary
and comments on possible applications.

\mathon
\section{$U(1)$-gauged $D\equ3$, $N\equ2$ supergravity}
\mathoff

\label{sec_gauging}

We start this section by reviewing the lagrangian and transformations
rules for a non-linear sigma model coupled to $N=2$ supergravity. This
mainly serves to set our notation and conventions; the reader is
referred to~\cite{dWToNi93} for full results and a detailed
discussion. Assuming an $U(1)$ isometry of the K\"ahler potential, we
apply the standard Noether procedure to obtain the general
$U(1)$-gauged $N=2$ supergravity. In four dimensions, analogous
supergravity theories have been studied
in~\cite{WitBag82,Bagg83,CFGP83}. We then evaluate the general
formulas in several examples.

\subsection{$D\equ3$, $N\equ2$ supergravity and K\"ahler geometry}

\label{ungauged}

The gravity multiplet of the ungauged $N\!=\!2$ supergravity in three
dimensions consists of a dreibein $e_\mu{}^a$ and two gravitini which
we assemble into one complex spinor $\psi_\mu$. The matter sector is
given by $\num$ copies of the $N\!=\!2$ scalar multiplet, each
consisting of $2$~real scalars and fermions. Again we use complex
notation $(\phi^\Ga, \Bphi{}^\BGa)$ and $(\Gl^\Ga, \BGl{}^\BGa)$ ($\Ga
= 1,\ldots \num$), respectively. The scalar fields parametrise a
K\"ahler manifold of real dimension $2p$, characterised by its
K\"ahler potential $K(\phi^\Ga, \Bphi{}^\BGa)$.

The $N\!=\!2$ locally supersymmetric lagrangian is given
by~\footnote{We use $2\times 2$ matrices $\Gg^a$ for the $SO(2,1)$
Dirac algebra, with $\Gg^a \Gg^b = \eta^{ab} -\varepsilon^{abc}
\Gg_c$; the charge conjugation matrix is $\Gg^0$. Our metric has
signature $(-++)$, and $\varepsilon^{012} =1$.  A convenient
representation is $\Gg^0 = i\sigma^3$, $\Gg^1 = -\sigma^2$ and $\Gg^2
= -\sigma^1$. We work in natural units. The sign of the Newton
gravitational coupling, which in three dimensions is not physically
fixed, is taken to be positive.}
\ba
\CL^S &=& 
\ft14eR -e\,G_{\Ga\BGa}(\phi,\Bphi)\,
\left(
\dd_\mu\phi^\Ga\,\dd^\mu\Bphi{}^\BGa 
+ \Blambda^\BGa \Gg^\mu D_\mu \lambda^\Ga
\right)
+\Ge^{\mu\nu\rho}\,\Bpsi_\mu D_\nu\psi_\rho
\non[1ex]
&&{}
+e\,G_{\Ga\BGa}(\phi,\Bphi)\,\left(
\Blambda^\BGa\Gg^\mu\Gg^\nu\psi_\mu\,\dd_\nu\phi^\Ga
+\Bpsi_\mu \Gg^\nu\Gg^\mu\lambda^\Ga\,\dd_\nu\Bphi{}^\BGa \right)
\;, \label{ungL}
\ea
up to terms quartic in the fermions. Here, $G_{\Ga\BGa}(\phi,\Bphi)$
is the K\"{a}hler metric $G_{\Ga\BGa}(\phi,\Bphi) = \dd_\Ga \dd_\BGa
K(\phi,\Bphi) $. The K\"{a}hler covariant derivatives acting on the
fermions are
\ba
D_\mu\psi_\nu &=& 
\nabla_\mu \psi_\nu -
\left( Q_\Ga\,\dd_\mu\phi^\Ga - 
Q_\BGa\,\dd_\mu\Bphi{}^\BGa \right)\psi_\nu \;,
\non
D_\mu\lambda^\Gb &=& 
\nabla_\mu \lambda^\Gb -
\left( Q_\Ga\,\dd_\mu\phi^\Ga - 
Q_\BGa\,\dd_\mu\Bphi{}^\BGa \right)\lambda^\Gb 
+ \GG^\Gb_{\Ga\Gg}\,\dd_\mu\phi^\Ga\lambda^\Gg \;,
\label{KahlD}
\ea
with spin-, $SO(2)-$, and K\"ahler- connections
\ba
\nabla_\mu &=& \dd_\mu + \ft14\,\Go_\mu{}^{ab}\Gg_{ab} \;,
\non
Q_\Ga(\phi,\Bphi) &=& \ft12\, \dd_\Ga K(\phi,\Bphi) \;,\qquad
Q_\BGa(\phi,\Bphi) = \ft12\, \dd_\BGa K(\phi,\Bphi) \;,
\non
\GG^\Ga_{\Gb\Gg}(\phi,\Bphi) &=& 
G^{\Ga\BGa}(\phi,\Bphi)\, \dd_\Gb G_{\Gg\BGa}(\phi,\Bphi) \;. 
\ea
Up to cubic terms, the supersymmetry transformations which leave the
lagrangian~\Ref{ungL} invariant are  given by
\ba
\Gd e_\mu{}^a &=& \Beps\Gg^a\psi_\mu - \Bpsi_\mu\Gg^a\Geps \;, \non
\Gd \psi_\mu &=& D_\mu\Geps ~=~ 
\nabla_\mu \Geps -
\left( Q_\Ga\,\dd_\mu\phi^\Ga - 
Q_\BGa\,\dd_\mu\Bphi{}^\BGa \right)\Geps \;,\non
\Gd \phi^\Ga &=& \Beps\lambda^\Ga \;, \non
\Gd \lambda^\Ga &=& \dd_\mu\phi^\Ga \,\Gg^\mu\Geps \;,
\ea
where   $\Geps$ is a complex spinor.

\subsection{K\"ahler transformations and isometries}

\label{symmK}

The sigma model geometry of the previous subsection is clearly
invariant under the K\"ahler transformations
\be
K(\phi,\Bphi) \ra K(\phi,\Bphi) + F(\phi) + \BF(\Bphi) \;,
\label{Kahl} 
\ee
while the potential $Q_\Ga$ transforms as
\be
Q_\Ga \ra  Q_\Ga + \ft12\,\dd_\Ga F(\phi) \;.
\ee
As in higher dimensions \cite{WitBag82,Bagg83}, the
lagrangian~(\ref{ungL}) is invariant under these transformations,
provided the fermionic fields simultaneously transform as
\ba
\psi_\mu &\ra& \exp(i\Im F)\,\psi_\mu\;,
\qquad
\lambda ~\ra~ \exp(i\Im F)\,\lambda \,.
\la{fermrot}
\ea
Moreover, we assume that the K\"ahler potential possesses an $U(1)$
isometry generated by the vector field
\be
\xi = \I\left( \phi^\Ga\dd_\Ga - \Bphi{}^\BGa\dd_\BGa \right)  \;,
\ee
i.e., $\xi$ generates a K\"ahler transformation
\be
\CL^S_\xi K  ~=~ \I\left( \phi^\Ga\dd_\Ga - 
\Bphi{}^\BGa\dd_\BGa \right) K ~=~ 
f(\phi) +\Bf (\Bphi) \;.
\label{LieKf}
\ee
The lagrangian \Ref{ungL} is then invariant under the global $U(1)$
\ba
\Gd \phi^\Ga &=& \I q\,\phi^\Ga \;,
\non
\Gd \psi_\mu &=& \I q\, \Im f\,\psi_\mu \; ,
\non
\Gd \lambda^\Ga &=& \I q\, (1+\Im f)\,\lambda^\Ga \;,
\label{U1f}
\ea
where the functional parameter $f$ is defined in terms of the
K\"{a}hler potential $K$ by equation \Ref{LieKf}. Observe that
(\ref{LieKf}) determines the holomorphic function $f$ only up to a
constant imaginary shift
\be
f \ra f+ \I c \;,
\label{cshifts}
\ee
with $c$ real.

\subsection{The gauged lagrangian and transformation rules}

\label{gauged}

We proceed to gauge the abelian K\"ahler isometry~(\ref{U1f}) of
the locally supersymmetric Lagrangian~(\ref{ungL}).  As in the maximal
theories \cite{NicSam01,NicSam01a}, this requires the coupling to a
Chern-Simons gauge potential $A_\mu$ with abelian field strength
$F_{\mu\nu} = 2\partial_{[\mu} A_{\nu ]}$ via
\be
\CL^{CS} = 
-\frac{1}{8}\, g \varepsilon^{\mu\nu\rho} A_\mu F_{\nu\rho} \; .
\label{LCS}
\ee
Note that this term is topological and hence does not introduce new
propagating degrees of freedom in the gauged theory.  The derivatives
in~(\ref{KahlD}) are gauge covariantized according to
\ba
\CD_\mu \phi^\Ga &\equiv& \dd_\mu \phi^\Ga + 
\I g\, A_\mu \phi^\Ga \;, \label{defD}
\\
\CD_\mu \psi_\nu &\equiv& 
\nabla_\mu \psi_\nu -
\left( Q_\Ga\,\CD_\mu\phi^\Ga - 
Q_\BGa\,\CD_\mu\Bphi{}^\BGa \right)\psi_\nu 
-\I g\,(\Im f)\,A_\mu \psi_\nu
\non
&=& 
\nabla_\mu \psi_\nu -
\left( Q_\Ga\,\dd_\mu\phi^\Ga - 
Q_\BGa\,\dd_\mu\Bphi{}^\BGa \right)\psi_\nu 
+\I g\,C A_\mu \psi_\nu \;,
\non
\CD_\mu \lambda^\Ga &\equiv& 
\nabla_\mu \lambda^\Gb -
\left( Q_\Ga\,\CD_\mu\phi^\Ga - 
Q_\BGa\,\CD_\mu\Bphi{}^\BGa \right)\lambda^\Gb 
+ \GG^\Gb_{\Ga\Gg}\,\CD_\mu\phi^\Ga\lambda^\Gg 
+\I g\,(1+\Im f)\,A_\mu \lambda^\Ga \;,
\nn
\ea
where we have defined the real K\"ahler invariant function
\be
C ~=~ -Q_\Ga\,\phi^\Ga - Q_\BGa\,\Bphi{}^\BGa 
- \ft12\,\I\, (f -\Bf ) \;.
\ee 
The latter transforms under the constant shifts~(\ref{cshifts}), viz.\
as
\be
C \ra C + c \;.  \label{cterm}
\ee
As will be seen shortly, this function is intimately related to the
superpotential of the gauged theory. The freedom in picking a constant
$c$ in~(\ref{cterm}) can then be related to the existence of the
Fayet-Iliopoulos term.

As usual, the $g$-dependent terms introduced above give rise to extra
terms in the supersymmetry variation of \Ref{ungL} according to
\ba
{}[\CD_\mu ,\CD_\nu ] \,\psi_\rho &=& 
\ft14\,R_{\mu\nu}{}^{ab}\,\Gg_{ab}\,\psi_\rho
+ 2 G_{\Ga\BGa}\,\CD_{[\mu}\phi^\Ga\,\CD_{\mu]}\Bphi^\BGa\,\psi_\rho 
+ \I g \,C F_{\mu\nu}\,\psi_\rho \; .
\ea
In order for these terms to vanish, extra Yukawa-type bilinear
fermionic terms and a scalar potential $V(\phi,\Bphi)$ must be added:
\ba
\CL^Y &=& -eg\,\Bpsi_\mu\,\Gg^{\mu\nu}\,\psi_\nu\,B -
eg\,\Blambda^\BGa\lambda^\Ga\,S_{\Ga\BGa} +
eg\,C\,(\Bpsi_\mu\Gg^\mu\lambda^\Ga\,S_\Ga-
\Blambda^\BGa\Gg^\mu\psi_\mu\,S_\BGa ) \;,
\non
\CL^V &=& e g^2 V \;.
\label{Yu}
\ea 
Here, the functionals $B(\phi, \Bphi )$ and $V(\phi, \Bphi )$ are real
scalars, while the vector functionals $S_\Ga (\phi, \Bphi )$ and the
tensor functionals $S_{\Ga \BGb}(\phi,\Bphi )$ are complex. Their
dependence on the scalars $\phi ,\Bphi$ will be specified below.

The supersymmetry variations are likewise modified by $g$-dependent
contributions:
\ba
\Gd e_\mu{}^a &=& \Beps\Gg^a\psi_\mu - \Bpsi_\mu\Gg^a\Geps \;,\non
\Gd \psi_\mu &=& \CD_\mu\Geps - g B \Gg_\mu \Geps  \;,\non
\Gd \phi^\Ga &=& \Beps\lambda^\Ga \;, \non
\Gd \lambda^\Ga &=& \CD_\mu\phi^\Ga \,\Gg^\mu\Geps 
-g G^{\Ga\BGa} C S_\BGa \,\Geps
\;,\non
\Gd A_\mu &=& -2\I(\Beps\psi_\mu-\Bpsi_\mu\Geps) \,C
+ 2\I\, G_{\Ga\BGa} \left( \phi^\Ga \Blambda^\BGa\Gg_\mu\Geps +
\Bphi{}^\BGa \Beps\Gg_\mu\lambda^\Ga \right) \;.
\la{susyG}
\ea
A straightforward calculation shows that --- modulo higher order
fermionic terms which presumably remain unchanged as in the maximal
theories of \cite{NicSam01,NicSam01a} --- the full Lagrangian
\be
\CL = \CL^S  +\CL^{CS} +\CL^{Y} + \CL^{V} ,
\la{gaugedL}
\ee
is invariant under \Ref{susyG} if the functionals $B, S_\Ga,
S_{\Ga\BGa}$ satisfy the following set of consistency relations:
\ba
\dd_\Ga B &=& -C \,S_\Ga\;,
\non
S_\Ga &=& 2 G_{\Ga\BGa} \Bphi{}^\BGa \;,
\non
D_\Gb S_\Ga &\equiv& \dd_\Gb S_\Ga - \GG^\Gg_{\Ga\Gb} S_\Gg ~=~ 0 \;,
\non
S_{\Ga\BGb} &=& B G_{\Ga\BGb} - S_\Ga S_\BGb + C\dd_\BGb S_\Ga ,
\label{constr}
\ea
while the potential is given by
\ba
V &=& 2 B^2 - G^{\Ga\BGa}\,C^2\,S_\Ga S_\BGa ~=~ 
2 B^2 - G^{\Ga\BGa}\,\dd_\Ga B \dd_\BGa B
\;.
\la{Vfinal}
\ea
It is easily checked that the system~(\ref{constr}) is consistent. The
general solution to these equations takes the form
\ba
B&=& C^2 + b \;, \non
S_\Ga &=& -2 \dd_\Ga C ~=~ 2\,G_{\Ga\BGa} \Bphi{}^\BGa \;,
\non
S_{\Ga\BGb} &=& 
B G_{\Ga\BGb} - S_\Ga S_\BGb -  2\,C \,\dd_\Ga \dd_\BGb C \;,
\label{sol0}
\ea
where $b$ is an arbitrary constant real parameter.

This completes the construction of a family of $N\!=\!2$
supersymmetric gauged lagrangians parametrized by a K\"ahler manifold
and two real numbers $b$ and $c$. In particular, the constant shifts
of $C$ by $c$ corresponds to the presence of a Fayet-Iliopoulos
term~\cite{Bagg83}. Note that $B=C^2\!+\!b $ is the superpotential for
$V$. In the limit $g\rightarrow 0$, one recovers the ungauged theory
\Ref{ungL}.

The gauged lagrangian \Ref{gaugedL} is still invariant under general
K\"ahler transformations \Ref{Kahl}, \Ref{fermrot}, under which the
functional parameter $f$ changes as
\be
f\ra f + \I \phi^\Ga \dd_\Ga F \;.  \la{residual}
\ee

\subsection{Some examples}

\label{special}

In this section, we consider some special cases of the abelian gauged
lagrangian constructed in the previous section. In particular, this
will reproduce and explain the form of the eighth order potential of
the abelian Chern-Simons Higgs model coupled to gravity, which was
previously derived in \cite{Valt92}--\cite{CCKK01} by Bogomol'nyi type
arguments. Moreover, we reproduce the gauged theories constructed in
\cite{DKSS00} and their generalization by Fayet-Iliopoulos terms.

In this paper, we will mainly be interested in the case $\num=1$,
i.e.\ of a single complex scalar field. Furthermore, we will restrict
ourselves to K\"ahler manifolds for which $f$ as defined in
\Ref{LieKf} is an imaginary constant $f\!=\!\I C_0$, i.e.\ $K$ is a
function of $R\!=\!|\phi|$ only. We then use the notations
\ba
\phi&\equiv&\phi^1 = R e^{i\phi}\;,\non
Q &\equiv& Q_{\overline{1}} ~=~  \frac{\phi}{4R}\,K' \;, \non
G &\equiv& G_{1\overline{1}} ~=~ \frac{1}{4R}\,(RK''+K') ,
\la{CQG}
\ea
while for the Yukawa tensors and the potential we find
\ba
B &=& C^2 + b \;,\qquad C ~=~ -\ft{1}{2}\,RK' + C_0 \;,
\non
S&\equiv&S_{\overline{1}} ~=~  \frac{\phi}{2R}\,(RK''+K') \;,\non
{\bf S}&\equiv&S_{1\overline{1}} ~=~ 
BG - |S|^2 -C\,C'' \;,\non
V &=&  2B^2 - 4GR^2C^2 \;.
\ea
Here $K'$, $K''$, etc.\ denote the derivatives of $K$ with respect to
$R$.

The simplest example in this class of models is the complex plane,
with K\"ahler potential $K(R)=R^2$. The above formulas then reduce to
\ba
G&=&1 
\;,\non
C&=&-R^2 + C_0 
\;,\non
V_0 &=& 
-4 R^2 (R^2- \eta^2)^2 + 
2 (R^4 - 2 R^2 \eta^2 + \phi_\infty^2 \eta^4)^2 \;,
\la{Higgs}
\ea
where we have set 
\ben
\eta^2 = C_0\;,
\qquad
\phi_\infty^2 = \frac{b+C_0^2}{C_0^2} \;.
\een
This precisely reproduces the eigth order potential of the abelian
Chern-Simons Higgs model coupled to gravity which was derived in
\cite{Valt92}--\cite{CCKK01} by requiring the existence
of Bogomol'nyi type equations. What we have shown here is that this
form of the potential is naturally explained by supersymmetry: it is
the unique potential which allows the embedding into an $N\!=\!2$
locally supersymmetric theory \Ref{gaugedL}. This holds for any choice
of parameters $\eta$ and $\phi_\infty$. In the following section, we
will see in more detail that the Bogomol'nyi type equations found in
\cite{Lond95} indeed descend from the supergravity Killing spinor
equations. 

For other examples we may consider the K\"{a}hler potentials
\be
K_\Ge (R)=\frac{\Ge}{a^2} \ln(1+\Ge R^2)  \;.
\ee
The cases $\Ge =+1$ and $\Ge =-1$ here correspond to the coset spaces
$S^2 = SU(2)/U(1)$ and $H^2 = SU(1,1)/U(1)$, respectively; the
constant $a$ denotes the characteristic curvature of these
manifolds. With the particular choice of parameters
\be
C_0 = \frac{\Ge}{2a^2}\;,\qquad b=0 \;,
\ee
the above formulas reduce to
\ba
G_\Ge &=& \frac{1}{a^2\,(1+\Ge R^2)^2}
\;,\non
C_\Ge &=& \frac{\Ge}{2a^2} \frac{1-\Ge R^2}{1+\Ge R^2} \;,\non
V_\Ge &=& 
\frac{(1-\Ge R^2)^2 \left((1-\Ge R^2)^2 - 8a^2 R^2 )\right) }
{8a^8 (1+\Ge R^2)^4}\;,
\la{DKSS}
\ea
which upon setting $g\!=\!4m a^4$ and rescaling $A_\mu\ra -\Ge g^{-1}
A_\mu$ precisely reproduces the two models studied in detail in
ref.~\cite{DKSS00}.\footnote{For a complete translation between the
notation of \cite{DKSS00} and that of the present paper, we note that
the matter fermions differ by rescaling with a vielbein living on the
K\"ahler manifold, which for $\num\!=\!1$ simply reduces to
$\sqrt{G}$; cf.~\cite{dWToNi93} for further details.}  Likewise, for
higher $\num$ and for particular choices of the K\"ahler manifold,
\Ref{gaugedL} reproduces the $CP^{\num}$ and $CH^{\num}$ models of
\cite{DKSS00}.

Our general construction furthermore yields a straightforward
generalization of \Ref{DKSS} by introducing a Fayet-Iliopoulos term,
i.e.~leaving $C_0$ as a free parameter
\be
C_0=\frac{\eta^2}{a^2 (1+\Ge \eta^2)} \;,
\ee
we find
\ba
G_\Ge &=& \frac{1}{a^2\,(1+\Ge R^2)^2}
\;,\non
C_\Ge &=& \frac{\eta^2-R^2}{a^2(1+\Ge R^2)(1+\Ge \eta^2)} \;,\non
V_\Ge &=& \frac{2(R^2-\eta^2)^2 
\left( (R^2-\eta^2)^2 - 2 a^2 R^2\,(1+\Ge \eta^2)^2 \right)}
{a^8 (1+\Ge R^2)^4 (1+\Ge \eta^2)^4} \;.
\la{DKSSgen}
\ea
These potentials have Minkowski vacua at $R=\pm \eta$. The particular
choices in \Ref{DKSS} correspond to $\eta^2=\Ge$. As an illustration,
in Figures~\ref{pot0} and \ref{potG} we depict the potentials of
\Ref{Higgs} and \Ref{DKSSgen} for the particular values of parameters
$g\!=\!a\!=\!1$, $\eta\!=\!1/4$. Their behaviour for $0<R<\eta$ is of
similar form, showing a $U(1)$-symmetric AdS vacuum at $R\equ0$ and
symmetry breaking Minkowski vacua at $R=\pm \eta$. Their global
behaviour however differs drastically with $V_+$ being bounded whereas
$V_0$ and $V_-$ become singular at infinite and finite $R$,
respectively. We will see in the next section that all these
potentials support supersymmetric vortex solutions interpolating
between $R=0$ and $R=\eta$.

\begin{figure}[htbp]
  \begin{center}
\epsfxsize=125mm
\epsfysize=60mm
\epsfbox{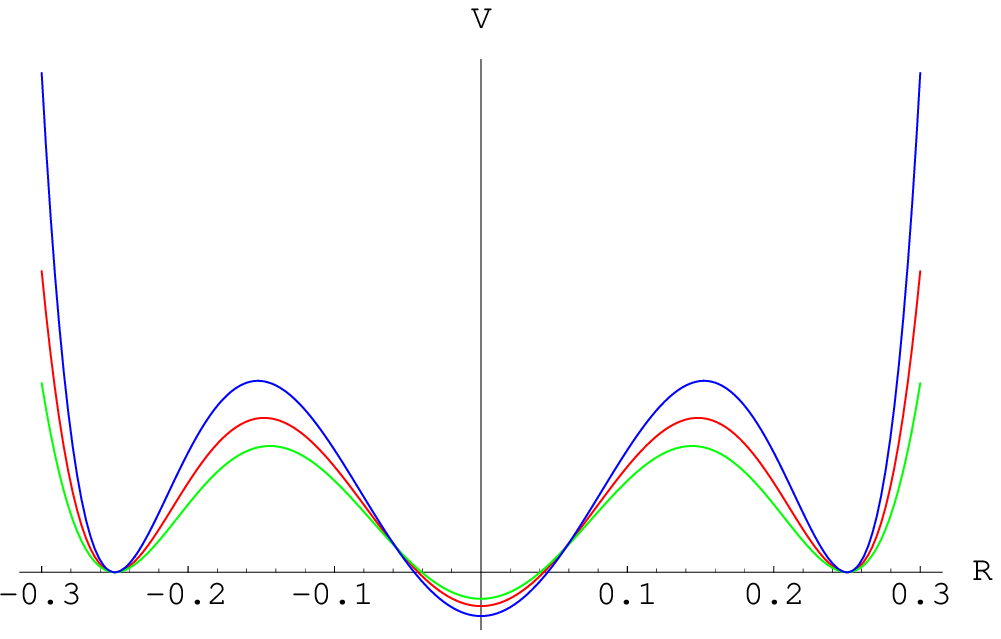}
  \caption{{\small Scalar potential $V$ for the models \Ref{Higgs},
\Ref{DKSSgen}: vacua at $R\equ0$, $R\equ\eta$.}}   
\la{pot0}
  \end{center}
\end{figure}
\begin{figure}[htbp]
  \begin{center}
\epsfxsize=125mm
\epsfysize=60mm
\epsfbox{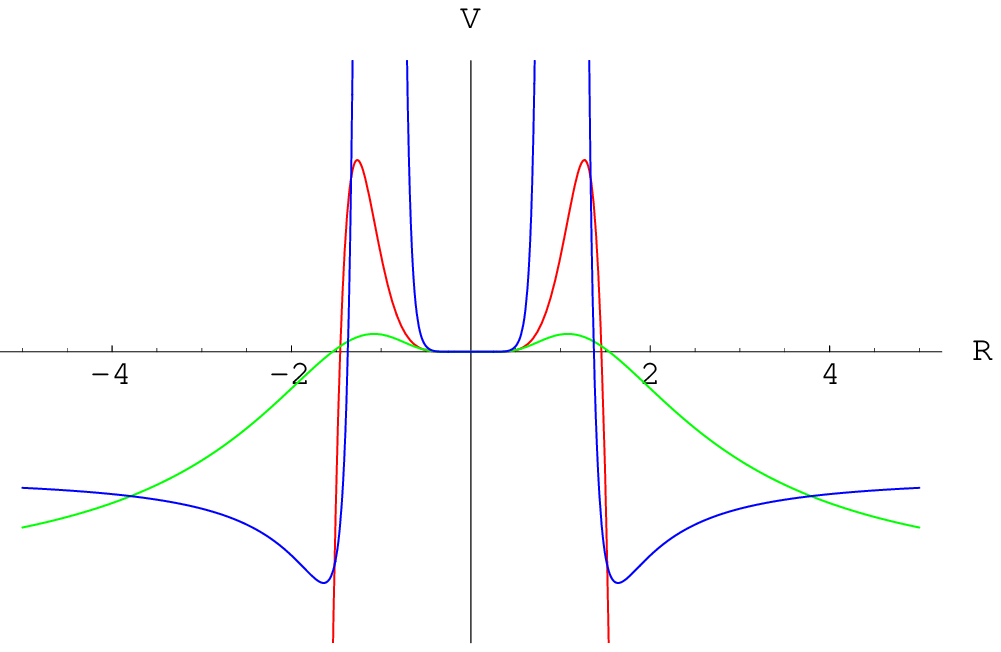}
  \caption{{\small Scalar potential $V$ for the models \Ref{Higgs},
\Ref{DKSSgen}: global behaviour.}} 
\la{potG}
  \end{center}
\end{figure}

\section{Supersymmetric vortices}

\label{sec_susy}

In this section we derive a set of first order differential equations
for rotationally symmetric, supersymmetric field configurations and
show that they admit regular vortex solutions. We restrict to models
with a single scalar field $p\equ1$ and a constant value of $f$ in
\Ref{LieKf}, i.e.\ to K\"ahler potentials depending on the absolute
value $|\phi|$ only. It would be interesting to generalize this
construction to K\"ahler potentials with an arbitrary holomorphic
$f=f(\phi)$, however this necessitates a more general ansatz for the
$\Gth$ dependence of the fields than will be considered here. For the
complex scalar $\phi$, we make the following time independent and
rotationally symmetric ansatz:
\be
\phi = R( r)\, e^{-\I n \Gth} \;.
\la{AnP}
\ee
For the vector field $A_\mu$ we choose the gauge in which
\be
A_r=0
\;,\quad
A_\Gth= P( r)+\frac{n}{g}
\;,\quad
A_t= W( r) \; .
\la{AnA}
\ee
Together with the boundary conditions at the origin and at infinity
discussed in detail in section \Ref{sectasympt} below, \Ref{AnP} and
\Ref{AnA} constitute our Ansatz for the static and rotationally
symmetric $|n|$-vortex.

For the three-dimensional metric, we take the (stationary and
rotationally symmetric) parametrization
\be
ds^2 ~=~ -dt^2 + 2\ell (r) \shc (r) \,dt d\Gth + 
\ell^2 (r) d\Gth^2 + dr^2 \;.
\la{AnM}
\ee
The Chern-Simons term in the lagrangian induces the first order
duality equations
\be
\Ge^{\mu\nu\rho}F_{\nu\rho} ~= ~ 
8\I eG\,  (\Bphi D^\mu \phi - \phi D^\mu \Bphi )  \;,
\la{duality}
\ee
relating the vector and the scalar field. With the ansatz
\Ref{AnP}--\Ref{AnM}, two of these equations take the form
\ba
\ell \cosh\chi \; \dd_r W &=& 
-4g\, G R^2  (P+\ell \sinh\chi  W)  \;, \non
\ell \cosh\chi \; \dd_r P &=& 
-4g\, G R^2 (\ell^2  W-\ell\sinh\chi P) \;,
\la{WpPp}
\ea
while the third one ($\mu\equ r$) is identically satisfied.

\subsection{Killing spinors}

For supersymmetric vortices we seek solutions to the Killing spinor
equations
\be
\Gd_\Geps\psi_\mu \stackrel{!}{=} 0 \;,\qquad 
\Gd_\Geps \Gl \stackrel{!}{=} 0 \; . 
\label{KS41}
\ee
The transformation of the matter fermions is given by
\be
\Gd_\Geps \Gl = \left( \CD_\mu \phi \Gg^\mu - 2g C\phi \right) 
 \Geps ~\stackrel{!}{=}~0 ,
\la{trafer}
\ee
together with the complex conjugate equation. The existence of
nontrivial solutions to this equation implies that
\be
( 2gC\phi )^2   = D_\mu \phi D^\mu \phi \;,
\non
\ee
which yields 
\be
4g^2 C^2  ~=~
g^2 W^2 + (\dd_r R/R)^2 -\frac{g^2}{\cosh^2 \chi} 
\left( W\sinh\chi +\frac{P}{\ell} \right)^2 \;.
\nn
\ee
This equation is solved e.g.\  by setting
\be
W = -2C   \;, \qquad\quad
\frac{\dd_r R}{R} = 
-\frac{g}{\cosh\chi}  \left(W\sinh\chi + \frac{P}{\ell}\right)  \;.
\la{WRpA}
\ee
Note that \Ref{CQG} implies the relation $\dd_R C = -2R G$. Utilizing
this, the ansatz \Ref{WRpA} is shown to be compatible with the duality
equations \Ref{WpPp}. Substituting \Ref{WRpA} back into \Ref{trafer},
the latter equation can be factorized as
\be
\left( 2\ell\, C-\I P\,\Gg^1\right)
\left( \cosh\chi + \I\Gg^0 -\sinh\chi\,\Gg^2 \right)
\Geps ~=~ 0 \; .
\la{proj1}
\ee
It is straightforward to check that the solution to this projector
equation is given by
\be
\Geps = \left(\cosh(\ft12\chi) + \sinh(\ft12\chi)\,\Gg^2 \right) 
\Geps_0 \;,
\qquad \mbox{with}\quad 
\left(1+\I \Gg^0 \right) \Geps_0 ~=~ 0 \;.
\la{eps}
\ee
It remains to study the first equation of \Ref{KS41}, i.e.\ the
requirement of vanishing transformation of the gravitini, which reads
explicitly
\be
\left[ \dd_\mu + \ft14\,\Go_\mu{}^{ab}\,\Gg_{ab} -
 (\BQ \dd_\mu \phi -Q \dd_\mu \Bphi ) +ig A_\mu C
-gB \gamma_\mu \right]  \Geps ~\stackrel{!}{=}~ 0 .
\la{KS}
\ee
For $\mu=\Gth$ this yields
\be
\dd_\Gth\, \Geps = 
-\left( \I g PC+\frac{n}{2}(f-\Bf )\right) \, \Geps 
- \frac{1}{2}\,\ell'\Gg^0 \,\Geps -\frac{1}{4} 
\left(\ell\chi'-\ell'\tanh\chi -4g 
B\ell \right)
\Gg^1 \,\Geps .
\la{ks1}
\ee
Utilizing the projection \Ref{proj1} to eliminate $\Gg^1 \,\Geps$,
this can be equivalently written as
\ba
\dd_\Gth\, \Geps &=& -\I
\left( g PC+\ft{1}{2}\,nC_0\right) \, \Geps 
+
\frac{\I}{4\sinh\chi} 
\left(\ell\chi'-\ell'\tanh\chi-4g B\ell  \right)
\Geps
\non
&&{} -
\ft{1}{4} \coth\chi
\left(\ell\chi'+\ell'\tanh\chi-4gB\ell \right)
\Gg^0\Geps .
\la{ks2}
\ea
The term in $\Gg^0\Geps$ must vanish separately, which gives an
equation for the metric coefficients, viz.\
\be
\ell\chi'+\ell'\tanh\chi = 4gB\ell \;.
\la{m1}
\ee
Substituting this back into~\Ref{ks2}, we find
\ba
\dd_\Gth\, \Geps &=& -\I
\left( g PC+\ft{1}{2}\,nC_0 
+\frac{\ell'}{2\cosh\chi} \right)\; \Geps \;,
\nn
\ea
which is solved by separation of variables such that
\be
\ell' = -(2gPC + nC_0 + 2k)\, \cosh\chi \;,
\qquad
\dd_\Gth\,\Geps = \I k \Geps \;,
\la{m2}
\ee
where $k$ is an arbitrary real parameter. The Killing spinor equation
\Ref{KS} for $\mu=t$ may be treated similarly. Separating variables
and utilizing \Ref{m1} and the projection \Ref{proj1}, equation
\Ref{KS} reduces to
\be
W = -\frac{2B}{C} -\frac{k'}{gC} \;, \qquad 
\partial_t \Geps =ik'\Geps \;,
\ee
which coincides with \Ref{WRpA} above provided that $k' = -2gb$ and
thus determines the time dependence of the spinor.
Finally there remains $\mu=r$, for which \Ref{KS} takes the form
\ben
\dd_r \Geps ~=~ \ft12 \chi' \Gg^2 \,\Geps \;.
\een
This equation is indeed satisfied by the ansatz \Ref{eps}, provided
$\partial_r \Geps_0 = 0\,$. Together with \Ref{eps} and \Ref{m2}, this
implies that the Killing spinor is given by
\be
\Geps(t,\theta,r) ~=~ 
e^{\I k \theta}e^{-2igbt} 
\left(\cosh(\ft12\chi) +\sinh(\ft12\chi)\,\Gg^2 \right) 
\Geps_0 \;,
\qquad 
\left(1\!+\!\I \Gg^0 \right) \Geps_0 ~=~ 0 \;.
\la{spinor}
\ee
The field configurations which solve eqs.~(\ref{WRpA}), \Ref{m1}, and
\Ref{m2} preserve half of the space-time supersymmetry; the
corresponding covariantly Killing spinor is given by \Ref{spinor}.

\subsection{The Bogomol'nyi equations}

To summarize, we have shown that with the ansatz \Ref{AnP}--\Ref{AnM},
the duality and the Killing spinor equations reduce to the following
set of first order differential equations
\ba
\dd_rR &=& g R
\left(2  \thc \, C  - \frac{1}{\ell\chc}\,P\right) \;, 
\non[.5ex]
\dd_r P &=& 4g\,GR^2 
\left( \frac{2\ell} {\chc}\,C  +\thc \, P \right) \;,
\non[1ex]
\dd_r \ell &=& (\ell_\infty-2g PC)\,\chc \;,
\non[1.5ex]
\dd_r (\ell\shc) &=& 4g\,B\,\ell\chc\,  \;,
\la{dgl}
\ea
where we have defined the constant
\be
\ell_\infty \equiv -2 \left( k + nC_0 \right) \;.
\la{l8}
\ee
Moreover, straightforward (albeit tedious) computation shows that
every solution to \Ref{dgl} indeed gives a solution to the full set of
field equations derived from \Ref{gaugedL}. For the abelian
Chern-Simons Higgs model \Ref{Higgs}, these equations reduce to the
set of differential equations derived in \cite{Lond95}. In the next
section we shall show that for given $n$, these equations admit a
unique vortex solution with regular asymptotics.

\subsection{Asymptotics of the Bogomol'nyi equations}

\label{sectasympt}

We are mainly interested in topologically stable, finite energy vortex
solutions. In particular, we expect the scalar field $R$ to run from
the symmetric AdS vacuum at $r=0$, $R=0$ into a Minkowski vacuum at
$R=\eta$ for $r\ra\infty$. Hence, these solutions cannot be
continuously deformed into the vacuum solution.

More precisely, around the origin $r=0$ we assume the following
asymptotic AdS behaviour of the metric
\be
\ell=r+  \CO(r)  \;,\quad \chi= m\, r +  \CO(r^2)  \;,\qquad
\la{asy0}
\ee
where the constant $m$ in our conventions is given by
$m=4g\,(C_0^2\!+\!b)$ and gives the inverse AdS radius of the
metric. Regularity of the scalar and gauge field around $r=0$ then
requires (we assume $n>0$, $g>0$)
\be
R=R_0 \, r^{n}  + \CO(r^{n+1})  \;,\qquad 
P=-\frac{n}{g} +  \CO(r^{2n})  \;,
\la{asy0a}
\ee
and fixes $k=-1/2$ in \Ref{l8}. The constant $R_0>0$ is the only free
parameter in the asymptotics around zero. It will be fixed by
demanding regularity of the solution at $r\ra\infty$. For $r\ra\infty$
we assume the following behaviour of the matter fields and metric:
\ba
R(r)&\rightarrow & \eta>0 \;,\qquad
P(r)~\rightarrow ~ 0 \;,
\non
\ell(r)&\ra& \ell_\infty \, r \;,\qquad
\chi(r)~\ra~ \frac{\chi_\infty}{r}  \;,
\la{asy8}
\ea
with $\ell_\infty=1\!-\!2nC_0$ defined in \Ref{l8} above, and
$\chi_\infty = n^2/(2g\ell_\infty^2)$\,. Closer inspection of the
differential equations~\Ref{dgl} shows that demanding regular
asymptotics at infinity leaves one free constant which appears in
subleading order in $R$, cf.~\Ref{Rasy8} below. Asymptotically, the
metric \Ref{asy8} describes a locally flat space with deficit angle
\be
\Gd= 4n\pi C_0 \;,
\ee
or, more precisely a particle with mass $M=2nC_0$ and spin
$J=n^2/(2g)$, \cite{DeJaHo84,Line90,Lond95}. A well defined conical
geometry at radial infinity requires the upper bound
\be
n < \frac{1}{2C_0}  \;,
\la{M<}
\ee
for the vortex number $n$. The values $2nC_0\!=\!1$ and $nC_0\!=\!1$
correspond to cylindrical and spherical asymptotic geometry,
respectively \cite{DeJaHo84}. Since \Ref{asy8} requires the function
$B$ as well as the potential $V$ to vanish at radial infinity, i.e.\
$B(\eta)=V(\eta)=0$, together we find that
\be
C(R\!=\!\eta)~=~ 0\;, \qquad b=0 \; ,
\la{constants}
\ee
which fixes the constants $b$ and $C_0$. Recalling that $C'=-2RG$ and
that $G$ remains positive to ensure a nondegenerate kinetic scalar
term, this in particular implies that $C_0>0$. In turn,
\Ref{constants} already implies that $V'(\eta)=0$, i.e.\ at radial
infinity the scalar field runs into a Minkowski vacuum of the
potential.

We seek a solution of the system of differential equations \Ref{dgl}
which interpolates between the proper asymptotics \Ref{asy0},
\Ref{asy0a} for $r\!\ra\!0$ and \Ref{asy8} for $r\!\ra\!\infty$,
respectively. As we have seen, there is precisely one free parameter
in the asymptotic expansion around $r\!=\!0$. It is a nontrivial
problem whether, by properly choosing this parameter $R_0$, one may
find a set of functions $\{R, P, \ell, \chi\}$ which obey regular
asymptotics \Ref{asy8} at radial infinity also. In the remaining part
of this section, we shall show that this is indeed the case.

We start our analysis of the differential equations \Ref{dgl} with
some observations. Consider the quantity
\be
Z \equiv P^2 
-4\ell \sinh\chi \left( CP- \frac{\ell_\infty}{2g}  \right)
-4 \ell^2 B \;.
\la{iom}
\ee
{}From the differential equations \Ref{dgl}, one may verify that $Z$
is an integral of motion, i.e.\ $\dd_r Z=0$, and hence reduces the
number of unknown functions to three. We further observe that
\Ref{dgl} imply the following second order equation for the scalar
field:
\ba
\frac{\dd}{\dd r} 
\left( \frac{\ell\chc}{R}\,\frac{\dd R}{\dd r} \right)
&=&
8g^2\,C\, ( B - G R^2) \:\ell\chc \;.
\la{sec}
\ea

To analyze the existence of solutions with proper asymptotics at $r\to
0$ and $r\to \infty$, we introduce the new radial variable
\be
\Tr \equiv  \ell\chc \, R^{-2C_0} \, e^K  \;.
\la{rt}
\ee
The system of equations \Ref{dgl} then yields the following simple
radial equation:
\be
\frac{\dd \ln \Tr}{\dd r} ~=~ \frac{\ell_\infty}{\ell\chc} \;.
\la{drt}
\ee
Recall  that the bound for a regular asymptotically conical geometry
\Ref{M<} implies that $0<\ell_\infty<1$ so that \Ref{rt} is indeed a
well-defined coordinate transformation.~\footnote{In contrast,
assuming asymptotically cylindrical geometry corresponds to
$\ell_\infty=0$, in which case $\Tr$ is not a well defined coordinate
but rather a constant. More precisely, in this case \Ref{drt} may be
integrated to
\ben
\ell\chc = (R/R_0)^{1/n}e^{K(0)-K(R)} \;.
\een
The first equation of \Ref{dgl} together with \Ref{iom} then reduces
to a decoupled first order differential equation for $R$
\ben
\dd_r R ~=~ 
R \sqrt{ n^2 (R/R_0)^{-2/n}e^{2(K(R)-K(0))} + 4g^2C^2} \;, 
\een
which shows that its solution $R$ necessarily diverges at radial
infinity. With the ansatz \Ref{AnP}--\Ref{AnM}, there are hence no
regular solutions with asymptotically cylindrical geometry in these
models. This is in agreement with the discussion in \cite{CCKK01} for
the abelian Higgs model.} According to \Ref{asy0},
\Ref{asy8}, the new radial variable $\Tr$ has the asymptotic behaviour
\be
\Tr \ra R_0^{-2C_0}\cdot r^{\ell_\infty} \quad\mbox{as}\quad r\ra 0 \;,
\qquad
\Tr \ra {\ell_\infty}  \eta^{2C_0}e^{K(\eta)}\cdot r
\quad\mbox{as}\quad r\ra \infty \;.
\ee
The metric element \Ref{AnM} becomes
\be
ds^2 ~=~ - (dt - \ell\shc\,d\Gth)^2   + 
(\ell \chc)^2\: \frac{dz\,d\overline{z}}{z\overline{z}}  \;,
\qquad\mbox{with}\quad
z=\Tr^{1/\ell_\infty} e^{i\theta} \;.
\ee
The key observation for our analysis however is the fact that, in
terms of $\Tr$, the second order differential equation \Ref{sec}
completely decouples from the metric functions and takes the form
\ba
\frac1{\Tr}\,\frac{\dd}{\dd\Tr} 
\left( \Tr \, \frac{\dd U}{\dd\Tr} \right) 
&=& -\frac{\dd V_{\rm eff}}{\dd \, U} \;,
\qquad
U=\ln R \;,
\la{secord}
\ea
with an effective potential 
\be
V_{\rm eff} ~=~ 
-\frac{ 2 g^2}{ \ell_\infty^2} \, e^{4C_0 U} \, e^{-2K} C^2  \;.
\la{Veff}
\ee
The effective potential $V_{\rm eff}$ is negative definite (as $R>0$)
and it vanishes at minus infinity and at $U\!=\!\ln\eta$, where it has
a local maximum since $C(\eta)\!=\!0$. For the particular examples
given in \Ref{Higgs} and \Ref{DKSSgen}, the effective potential is
depicted in Figure~\ref{veff}.
\bigskip
\bigskip

\begin{figure}[htbp]
  \begin{center}
\epsfxsize=110mm
\epsfysize=50mm
\epsfbox{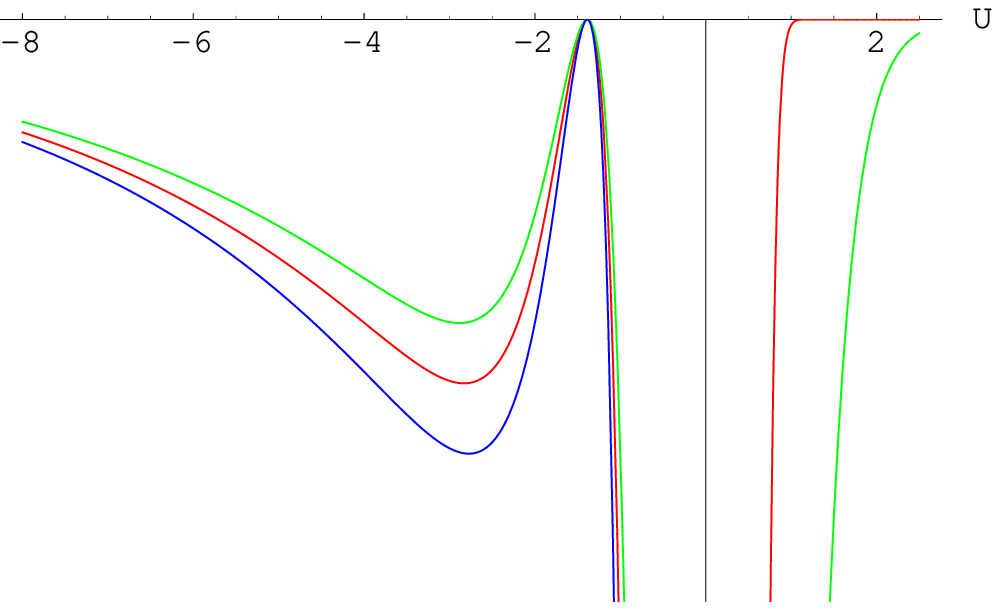}
  \caption{{\small Effective potential $V_{\rm eff}$ from \Ref{Veff}
for the  models \Ref{Higgs}, \Ref{DKSSgen}.}} 
 \la{veff}
  \end{center}
\end{figure}

Note that the gauge coupling constant $g$ may be absorbed by rescaling
$\Tr$. As in the case of nongravitating vortices with Higgs
potential~\cite{JaLeWe90}, equation \Ref{secord} can be approximated
by a Liouville equation for small $R$ and by a Bessel equation in the
vicinity of the Minkowski vacuum at $R=\eta$. More precisely, we find
in agreement with \Ref{asy0a}, \Ref{asy8} that
\be
R ~\approx~ \left( 2gC_0^2e^{-K(0)}\Tr \right)^{-\frac{1}{2C_0}}\:
\left( \Big( \frac{\Tr_0}{\Tr}\Big) ^\frac{1}{\ell_\infty}-
\Big(\frac{\Tr}{\Tr_0}\Big) ^\frac{1}{\ell_\infty}
\right)^{-\frac{1}{2C_0}}\;,
\qquad\mbox{for small $R$}\;,
\la{Rasy0}
\ee
with a constant $\Tr_0$ related to $R_0$ above, and that
\be
R ~\approx~ \eta + c_1 \,K_0\Big(\frac{2g}{\ell_\infty}\,
\eta^{2C_0+\frac12}e^{-K(\eta)}C'(\eta)\,
\Tr \Big) \;, \qquad\mbox{near $R=\eta$} ,
\la{Rasy8}
\ee
where $K_0$ denotes the Macdonald function and $c_1$ is a constant.
Starting either near $R=0$ or near $R=\eta$, the constants $\Tr_0$ and
$c_1$ are implicitly (and uniquely) fixed by requiring regular
asymptotics at the other end. Let us now discuss how this comes about
in slightly more detail.

The form of equation \Ref{secord} allows us to prove the existence of
a regular solution, which interpolates between the proper asymptotics
\Ref{asy0a} and \Ref{asy8}. This may be shown in a manner which is
reminiscent of the discussion in \cite{CCKK01}. The point is that the
second order differential equation \Ref{secord} can be thought of as
describing the damped motion of an auxiliary fictitious particle in
the effective potential \Ref{Veff}. There is a one-parameter family of
solutions which, at $\Tr\!=\!0$, start with the correct asymptotics,
viz.\
\be
U =  {\frac{n}{\ell_\infty}}\,\log\Tr + \log R_0 + \dots \; ;
\la{Ua}
\ee
these solutions are labeled by the parameter $R_0$ from
\Ref{asy0a}. Similarly to the analysis in~\cite{CCKK01}, one finds
that for small $R_0$ these solutions run into the local minimum of the
effective potential, whereas for large values of $R_0$ they go over
the hilltop of the potential at $U\!=\!\ln\eta$.  There is precisely
one value of $R_0$ for which the motion of this fictitious particle
ends at the local maximum $U\!=\!\ln\eta$. This corresponds to the
proper asymptotics \Ref{asy8} of the scalar field at radial infinity.

Having solved equation \Ref{secord} with the proper
asymptotics at both ends, it remains to restore the other fields $\{P,
\ell, \chi\}$ of the model. To this end, we note that the differential
equations \Ref{dgl} imply that
\be
\frac1{\Tr}\,\frac{\dd}{\dd\Tr} \left(\ell\shc\right) ~=~ 
-\frac{2\ell_\infty}{g}\,V_{\rm eff}  \;.
\la{ex0}
\ee
Using \Ref{secord}, this equation may be integrated to
\be
\ell\shc ~=~ \frac{\ell_\infty}{2g}\,
\left( \frac{n^2}{\ell_\infty^2} - 2\Tr^2\,V_{\rm eff} -
\Big(\Tr\,\frac{\dd U}{\dd\Tr}\Big)^2 
\right) \;,
\la{ex1}
\ee
and one easily verifies that this indeed obeys the correct asymptotics
\Ref{asy0}, \Ref{asy8}. The remaining metric function is obtained
from \Ref{rt}, which gives
\be
\ell\chc ~=~ \Tr\,R^{2C_0}\,e^{-K(R)} \;,
\la{ex2}
\ee
while the gauge field $P$ may be extracted from \Ref{dgl} as
\be
P ~=~ -\frac{\ell_\infty}{g}\,
\left( \Tr\,\frac{\dd U}{\dd\Tr} + 
2\Tr^2\,C\,V_{\rm eff} +
C \Big(\Tr\,\frac{\dd U}{\dd\Tr}\Big)^2 
- \frac{n^2}{\ell_\infty^2}\,C \right) \;.
\la{ex3}
\ee
All the fields have correct asymptotical behaviour \Ref{asy0},
\Ref{asy0a}, \Ref{asy8}, provided the function $R$ is the unique
solution of \Ref{secord} with proper asymptotics. The original radial
variable $r$ may finally be restored by integrating \Ref{drt},
\be
r ~=~ \frac1{\ell_\infty}\,\int_0^r   d\Tr\,R^{2C_0}\,e^{-K(R)}\;.
\la{rint}
\ee
To summarize, we have shown that the system of differential equations
\Ref{dgl} may be reduced to a single second order differential
equation \Ref{secord} which, for each vortex number $n$
satisfying~\Ref{M<}, admits a unique solution with proper asymptotics
at $r\!=\!0$ and $r\to \infty$. From this solution, the original
fields may be restored according to \Ref{ex1}--\Ref{ex3} and they have
the correct asymptotics.  This completes the construction of vortex
solutions in the general $U(1)$-gauged supergravity~\Ref{gaugedL}.

An obstruction to the existence of these vortex solutions may however
show up for certain compact K\"ahler manifolds. As has been observed
in the four-dimensional case in \cite{WitBag82}, global consistency of
the lagrangian~\Ref{gaugedL} requires the K\"ahler manifold to be a
Hodge manifold, which may imply a quantization of the gravitational
constant in units of the scalar self-coupling. For the compact $S^2$
model~\Ref{DKSS} for example, one finds the restriction
\be
2C_0 = \frac1{a^2} \in \bbZ \;,
\la{quant}
\ee
as has been explicitly verified in \cite{DKSS00}. This quantization
condition is obviously incompatible with the bound \Ref{M<}, and hence
reduces the possible values of the vortex number to $n\equ1$ and
$n\equ2$, corresponding to asymptotically cylindrical and spherical
geometry, respectively. As has been discussed above, these solutions
do not exist in this model. The absence of vortex solutions with
asymptotically conical geometry and the scalar fields living on a
compact target manifold has already been noted in four dimensions
in~\cite{ComGib88}.

\subsection{Examples}

Let us illustrate the analysis of the last section by computing some
numerical solutions to the specific models presented in
section~\ref{special} above. Since the abelian Chern-Simons Higgs
model coupled to gravity~\Ref{Higgs} and its vortex solutions have
already been extensively discussed in the literature, we will consider
the noncompact $H^2$ model with potential $V_-$ given in \Ref{DKSSgen}
above, whose vortex solutions have not yet been analyzed. In fact,
this model may be of special interest, since in the limit $g\ra0$ it
reduces to the theory obtained from dimensional reduction of pure
four-dimensional Einstein gravity. It is then tempting to speculate
about a possible higher-dimensional geometrical origin of these vortex
solutions. Recall that the choice of a strictly positive $C_0$ in this
model was essential for the existence of a Minkowski vacuum and hence
for the existence of the vortex solutions.

For the particular values $a\equ g\equ1$ and $\eta\equ\ft14$ of the
parameters, the scalar potential $V$ and effective potential $V_{\rm
eff}$ have been depicted in Figures~\ref{pot0}, \ref{potG} and
\ref{veff}, respectively. With these parameters, \Ref{M<} yields the
upper bound $n<8$ for the vortex number. For each $0<n<8$, the unique
solution $R(\Tr)$ to \Ref{secord} which has asymptotics~\Ref{Ua}
around $\Tr\!=\!0$ and remains regular at $\Tr\!=\!\infty$ may be
found numerically, by fine-tuning the unknown parameter $R_0$ by
hand. We should stress, however, that finding the regular solutions
with higher vortex numbers requires considerable numerical accuracy. A
nontrivial check is provided by inserting the solution obtained into
the effective potential $V_{\rm eff}$ and numerically integrating the
r.h.s.\ of equation~\Ref{ex0}, which should result in
\be
\int_0^\infty V_{\rm eff}(R(\Tr))\, \Tr\, d\Tr ~=~ 
-\frac{n^2}{4\ell_\infty^2} \;,
\ee
which is found upon integrating the l.h.s.\ of~\Ref{ex0} and
using~\Ref{asy8}. All our numerical solutions pass this check with
high precision (up to 0.005\%). The original radial variable $r$ is
finally obtained by numerically integrating \Ref{rint}. The resulting
functions $R(r)$ for all possible values $n=1, 2, \dots, 7$ of the
vortex number have been plotted in Figure~\ref{figR}, the value of $n$
increasing from left to right. The behaviour of the (normalized) gauge
field $P(r)$ in these solutions is given in Figure~\ref{figP}.
\bigskip
\bigskip

\begin{figure}[htbp]
  \begin{center}
\epsfxsize=110mm
\epsfysize=50mm
\epsfbox{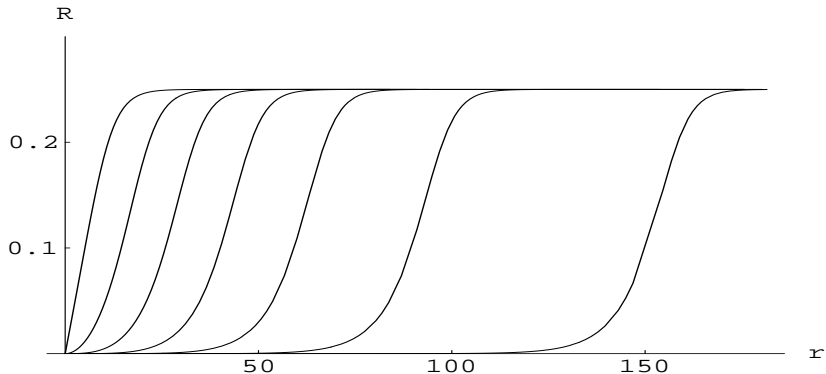}
  \caption{{\small Scalar field $R(r)$ for vortices in the $H^2$ model
with $n=1, 2, \dots,7$ .} }
 \la{figR}
  \end{center}
\end{figure}
\begin{figure}[htbp]
  \begin{center}
\epsfxsize=110mm
\epsfysize=50mm
\epsfbox{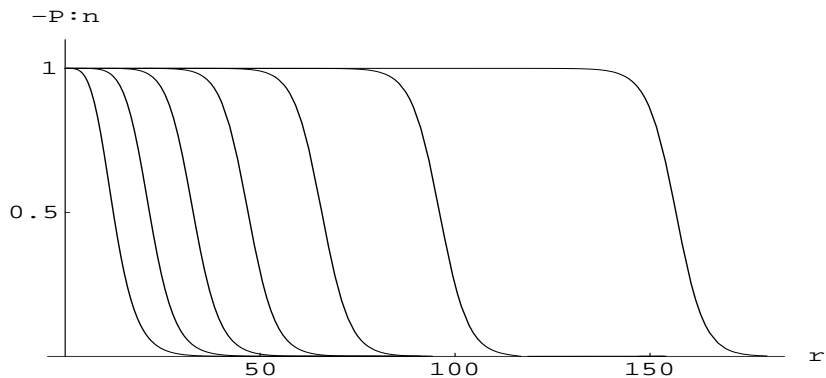}
  \caption{{\small Gauge field $-P(r)/n$ for vortices in the $H^2$
model with $n=1, 2, \dots,7$ .} }
 \la{figP}
  \end{center}
\end{figure}

\section{Summary and outlook}

\label{sectsumm}

We have constructed the general three-dimensional $N=2$ gauged
supergravities with abelian gauge group $U(1)$ and Chern-Simons
coupling of the vector fields. The supersymmetric lagrangian and
supersymmetry transformations rules were determined up to quartic and
cubic fermionic terms, respectively, but we do not expect the
structure of the models to be modified by the higher order fermionic
terms. The models are parametrized by a K\"{a}hler manifold and two
real numbers $b$ and $c$; $b$ shifts the superpotential for the scalar
potential (cf.\ eqs.~\Ref{Vfinal} and \Ref{sol0}), while a nonzero
value of $c$ corresponds to the presence of a Fayet-Iliopoulos term.
The gauged lagrangians \Ref{gaugedL} are residually invariant under
general K\"{a}hler transformations.

We have evaluated our general formulas for various particular examples
with a single complex scalar field $\phi$ and K\"{a}hler potential
$K\equ K(|\phi |)$. In the case of the complex plane, this reproduces
the eighth order polynomial potential~\Ref{Higgs} of the abelian
Chern-Simons Higgs model coupled to gravity which was previously
derived in~\cite{Valt92}--\cite{CCKK01} by requiring that the dynamics
be given by a system of first order differential equations. This
constitutes a natural explanation of these earlier results from local
$N=2$ supersymmetry. In particular, our construction provides the
embedding of the abelian Chern-Simons Higgs model into a supergravity
theory, allowing us to directly address the stability of the vortex
solutions preserving half of the supersymmetry. In flat space, the
self dual limit of the abelian Chern-Simons Higgs
model~\cite{JaLeWe90} and its particular sixth order potential have
similarly been derived from global $N=2$ supersymmetry
in~\cite{LeLeWe90}.

Likewise, for the K\"ahler manifolds $CP^\num$ and $CH^\num$, our
results reproduce the potentials of the gauged $N=2$ models studied
in~\cite{DKSS00}, together with their generalization by including a
Fayet-Iliopoulos term. The presence of this term allows for
symmetry-breaking Minkowski vacua of the potentials and hence for
vortex solutions.

Having constructed the general abelian gauged $N=2$ supergravity
theory, we turned to the construction of rotationally symmetric vortex
solutions preserving one half of the supersymmetry. Utilizing the
Ansatz~(\ref{AnP})-(\ref{AnM}) for the fields, the Killing spinor
equations \Ref{KS41} were shown to lead to the system of four first
order differential equations~\Ref{dgl} (which in particular reduces to
the system found in~\cite{Lond95} in the special case of the
Chern-Simons-Higgs model); it was also verified that this system
solves the full set of (second order) field equations of the
theory. Furthermore, we showed that this system of equations admits a
unique solution in which the norm of the scalar field $R$ runs from
the symmetric AdS vacuum of the potential at $r=0$, $R=0$ into a
symmetry breaking Minkowski vacuum at $r\to \infty$. This solution
represents a rotationally symmetric, finite energy, topologically
stable vortex solution. The essential ingredient for proving the
existence of this solution was the further reduction of the
system~\Ref{dgl} to the single second order differential equation
\Ref{secord}. The latter describes the motion of a (fictitious)
particle in the effective potential \Ref{Veff}. Its solution
determines the original fields (scalar, gauge field and components of
the metric) via \Ref{ex1}--\Ref{ex3}.

Our results provide further examples beyond that of
refs.~\cite{BeBeSt95,EdNuSc96} of the mechanism proposed
in~\cite{Witt95} for obtaining a vanishing cosmological constant
within a supersymmetric theory without phenomenologically unacceptable
Bose-Fermi degeneracies. In particular, the solution constructed here
is the first such example within a gauged supergravity with abelian
Chern-Simons gauge fields (rather than the usual Maxwell fields). The
covariantly constant spinors of our solutions exist by virtue of
essentially the same mechanism as pointed out in~\cite{BeBeSt95}.

Finally, let us mention some directions for further investigations. We
have constructed the general abelian gauged $N=2$ theory, but of
course it would be very interesting to obtain nonabelian gaugings and
to identify possible restrictions on the allowed gauge groups by
solving the consistency conditions imposed by local supersymmetry in
this case. One would expect such models to admit vortex type solutions
with several gauge fields turned on (i.e.\ nonabelian vortices) which
it would be interesting to construct explicitly, perhaps by acting
with some suitably adjusted solution generating transformations on the
abelian solutions constructed here. The possible relevance of vortex
solutions in the AdS/CFT correspondence has been addressed in
\cite{DeGhMa01}. We hope to report on these and related matters in the
near future.

\section*{Acknowledgements}

We would like to thank Jos\'{e} Edelstein for discussions and
collaboration in the initial stages of this work. We also thank the
Erwin Schr\"{o}dinger Institute and the organizers of the program
entitled {\em Mathematical Aspects of String Theory} for hospitality
and support in Vienna during its completion. The work of M.\ A.\ was
supported by the Swiss National Science Foundation under grant number
83EU-056178. This work was supported in part by EU contract
HPRN-CT-2000-00131.


\providecommand{\href}[2]{#2}
\begingroup\raggedright\endgroup

\end{document}